# APPLICATION SOFTWARE STRUCTURE ENABLES NIF OPERATIONS

Kirby W. Fong, Christopher M. Estes, John M. Fisher, Randy T. Shelton
LLNL, Livermore, CA 94550, USA

Abstract

The NIF Integrated Computer Control System (ICCS) application software uses a set of service frameworks that assures uniform behavior spanning the front-end processors (FEPs) and supervisor programs. This uniformity is visible both in the way each program employs shared services and in the flexibility it affords for attaching graphical user interfaces (GUIs). Uniformity of structure across applications is desired for the benefit of programmers who will be maintaining the many programs that constitute the ICCS. In this paper, the framework components that have the greatest impact on the application structure are discussed.

## 1 INTRODUCTION

The ICCS has extracted into framework software several capabilities needed by many applications [1]. Along with extensible program components, an object-oriented (OO) project needs to have patterns for combining the components into useful and uniform executable programs. In ICCS these patterns are expressed in several ways at varying degrees of formality. Generic procedures are instantiated to create main programs. Conventions that are reused for numerous user interfaces provide a common look and feel for different control panels. Recurring temporal patterns decouple threads of execution where several objects on separate computers interact. Idiomatic techniques make all exception handlers seem similar, with only the details of the reported exception, not the structure of the program, varying between examples. These patterns have been documented and are exploited at several levels of the ICCS implementation.

## 2 PROCESSES

All ICCS applications, whether FEPs, supervisors or servers, are instantiations of an Ada generic procedure (a generic is a code template). The templates supply the code for initializing the client side of several framework services, integrate the creation and initialization of application objects, and create the tasks (the threads of execution) that execute the methods of Common Object Request Broker Architecture (CORBA) objects when requests come in.

The local system manager is the generic main program that performs the client setup for message logging, event and alert propagation, and reservation frameworks. It is a proxy for the central system manager remote process.

The start-up process is a collaboration between the new main program instance, the central system manager, and the configuration server. The central system manager process mediates the starting of processes while assuring the temporal dependencies—first FEPs, then supervisors, finally GUIs—and continuously monitoring and reporting unscheduled process termination.

After instantiation of these framework services (which are common to all applications) is complete, the objects that carry out the business of the control system can be created by the configuration server. ICCS applications use the factory pattern [2] to postpone until start-up time binding the identity and number of objects that go into each application.

The generic main program requires two object factories, each CORBA objects. Each process's factories were custom built to implement the selection of CORBA objects called configurables, which it can create upon request; this selection of factory methods is the only feature that distinguishes the 18 different FEPs, or the numerous supervisor processes, from each other.

Each of the different supervisor processes and FEPs has two distinct factory objects that build, respectively, either local configurable objects or distributed configurable objects.

Local configurables are objects that are private to a process. In an FEP, these are the controllers, which model computer hardware, input/output boards, ports, channels, and protocols. Examples include commercially purchased drivers for video cameras and Ethernet protocol stacks. Controllers hide the implementation details of the hardware from the upper levels of the ICCS. The NIF Project uses a standardized common set of hardware controllers and has implemented a controller software that models their operational features. These controllers are reused in different FEP applications simply by adding a small block of code to an FEP's controller factory.

Distributed configurables are CORBA objects. Unlike local configurables, they export their interfaces to the rest of the system. The distributed configurables in an FEP are called devices. Devices typically model laser hardware and define the building blocks of the software system. Examples of devices include multi-axis actuators for optic alignment, which are implemented using as many as four hardware-specific controllers. Whereas controllers belong to the control point and computer hardware domains, devices belong in the laser design domain. Like controllers, devices can be reused across applications simply by adding a small block of code to an FEP's device factory.

ICCS application programs are all servers. When they start, they create and initialize their objects and then wait for clients to request service. During instantiation, the generic main program creates several (typically four) CORBA tasks that become the threads of execution that wait in an event loop. The number of these worker tasks determines the maximum number of different client interactions that can proceed concurrently. When all worker tasks are occupied, subsequent invocations are queued by the Ada runtime and proceed first-in, first-out.

## 3 SUPERVISORY INTERACTIONS

The Logical Control Unit (LCU) is the base class of supervisor objects that model the physical portions of the NIF above the device control point level. Supervisory applications partition their responsibilities among LCUs to reflect the grouping of the hardware.

The supervisory application architecture requires that LCUs frequently need to communicate with each other and also that GUIs are not tightly bound to particular application objects. The architectural style described by Taylor et al [3] led the LCU base class to have methods for both sending and receiving messages – the so-called observer pattern [2].

The left side of Figure 1 is an abbreviated class diagram in the Unified Modeling Language notation. The Director class has an update subscriber method that any publisher can call to deliver information to the subscriber. Thus, any derivative of Director, particularly the GUI object, can be a subscriber.

In addition, each LCU is a derivative of Director that adds a publishing capability. Both GUIs and LCUs subscribe to update messages from an arbitrary collection of publishers. These messages are constructed by a class called Data Mapper, which insulates the private data representation within an LCU from subscribers' needs. An LCU can have many data mappers, each capable of producing some piece of information, and clients can subscribe to any of the data mappers of an LCU.

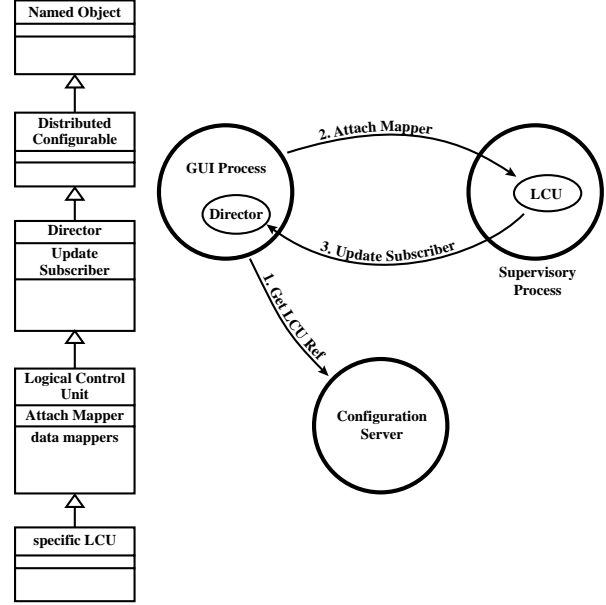

Figure 1: GUIs use application framework classes.

The right side of Figure 1 shows some typical GUI interactions. A client GUI consults the Configuration Framework to obtain a Ref that embodies CORBA connectivity to the desired LCU, then starts the subscription by making a call to the Attach Mapper method of the LCU. Because of this loose coupling between the GUIs and supervisors, it is feasible to bring up multiple instances of the same GUI, all subscribing to the same data mappers. This allows operators to bring up the same display at multiple terminals.

Subscribers and publishers are both CORBA objects and therefore can reside in separate processes. In fact, most LCUs are implemented as Ada objects in supervisory applications, while the GUIs implement their Directors in Java and exploit the substantial capabilities of the Java environment to render graphic windows.

## 4 STATUS PROPAGATION

The status of every device is observed by one or more supervisory objects and can be displayed on GUIs upon operator demand. The LCUs employ the status monitor framework to avoid network polling. The latency and precision of monitor reports are individually set by each observing LCU so each maintains a state that is an appropriately precise image of the device status. Only changes in a device state that are deemed significant by the subscriber are sent to the supervisory application.

The framework is built of components that are installed in both the FEPs and the LCUs. A supervisory application requests a device to begin status monitoring. The device engages the services of the status monitor, which then begins polling and sending messages about significant changes back to an object contained in the LCU. These updates cause the LCU to evolve its state to reflect the device's status and typically to publish new information about its evolving state using data mappers as described above.

## 5 GRAPHICAL USER INTERFACES

A GUI framework is provided to facilitate the development process. GUI framework classes provide a common set of mechanisms for instantiating graphical interfaces and directing them to connect to their devices. General mechanisms are provided for starting the control panel window, for connecting to the devices and LCUs to be controlled, and for uniform presentations such as help menus. All visual style elements, such as colors and fonts, are defined in the GUI frameworks and not in individual applications. Java classes have been built to implement both the general features of the framework and the specifics of numerous control interfaces needed for both supervisory and FEP controls.

Common behavior has been built that permits copies of the same graphics element to be used either for the same or for duplicate physical devices. GUI management applications (e.g., a broadview display or a device selector) can invoke control panels without knowing anything about the subordinate besides its class name and the identity of the device it controls.

## 6 CORBA CONNECTIONS

Regardless of the particular CORBA product being used, certain classes of error condition must be addressed in any distributed application. The ICCS software framework provides failure recovery mechanisms that allow application software to function in the presence of communication failures. Preliminary work has defined connection abstractions that define a uniform group of exceptions to which all clients are expected to respond uniformly.

Typically there are three situations in which communication failures can occur, corresponding to initial connection failure, failure of a previously successful connection, and failure in the middle of a two-way invocation. The ICCS framework supports various connection abstractions that clients can configure for different connection behaviors.

Initial connection failure can occur on the first invocation on a reference received from another source. A name service – the configuration framework – is used to provide CORBA references to client applications. ICCS addresses initial failure errors by allowing application programs to be configured to wait for services' presence before starting application objects.

Previously successful connection failure can occur when some object in the system fails after a successful connection has been established. The client side process does not know that something unusual has occurred until it tries to reuse a formerly reliable connection and receives a communication failure exception. This failure mode is ameliorated by a connection object that can be configured to ping the target service prior to each invocation. If the ping fails, the connection abstraction can execute recovery logic that will attempt to refresh the CORBA reference.

Failures in the middle of an invocation caused by rare process deadlocks, have proved especially difficult to diagnose. These are investigated by tracing tools. Future usage of network management may make this failure mode more amenable to detection.

## 7 SUMMARY

Experience has shown that construction of uniform, pattern-compliant application programs have several benefits. Validated frameworks are widely understood by the developer team so that implementation tasks can be shared. Code inspection can assure consistent behavior among programs that share frameworks. Enhancements in framework functionality are easily imported into all applications.

## REFERENCES

[1] R. Carey, et al., "Large-scale CORBA-distributed Software Framework for NIF Controls," ICALEPCS 2001.
[2] E. Gamma, R. Helm, R. Johnson, and J. Vlissides, "Design Patterns," Addison-Wesley Publishing Company, Reading, Massachusetts, 1995.
[3] R.N. Taylor et al., "A Component- and Message-Based Architectural Style for GUI Software," IEEE Transactions on Software Engineering, Vol. 2222, No. 6, June 1996, pp 390–406.